\newcommand{\mrad}{\ensuremath{\mathrm{mrad}}}

\newcommand{\mum}{\ensuremath{\mathrm{\mu m}}}

\newcommand{\mm}{\ensuremath{\mathrm{mm}}}
\newcommand{\GeVp}{\ensuremath{\mathrm{GeV/c}}}
\newcommand{\R}{\ensuremath{R}}
\newcommand{\rvar}{\ensuremath{r}}
\newcommand{\x}{\ensuremath{\mathrm{x}}}

\documentclass{elsart}
\usepackage{natbib}
\usepackage{graphicx}
\usepackage{glas}

\begin{document}

\begin{titlepage}{GLAS-PPE/2008-07}{10$^{\underline{\rm{th}}}$ July 2008}
\title{Performance of the LHCb Vertex Detector Alignment Algorithm determined with Beam Test Data}
\author{M.~Gersabeck$^{*a}$,S.~Viret$^{*a}$,C.~Parkes$^{*a}$ {\it et al.}\\
(the LHCb VELO group)
\\
$^a$ University of Glasgow, Glasgow, United Kingdom\\
{\it $^*$ corresponding authors}}

\begin{abstract}
LHCb is the dedicated heavy flavour experiment at the Large Hadron Collider at CERN. The partially assembled silicon vertex locator (VELO) of the LHCb experiment has been tested in a beam test. The data from this beam test have been used to determine the performance of the VELO alignment algorithm. The relative alignment of the two silicon sensors in a module and the relative alignment of the modules has been extracted. This alignment is shown to be accurate at a level of approximately $2~\mum$ and $0.1~\mrad$ for translations and rotations, respectively in the plane of the sensors. A single hit precision at normal track incidence of about $10~\mum$ is obtained for the sensors. The alignment of the system is shown to be stable at better than the $10~\mum$ level under air to vacuum pressure changes and mechanical movements of the assembled system.
\end{abstract}

\newpage
\end{titlepage}

\runauthor{Marco Gersabeck}
\begin{frontmatter}
\title{Performance of the LHCb Vertex Detector Alignment Algorithm determined with Beam Test Data}

\author{M.~Gersabeck$^{*a}$,S.~Viret$^{*a}$,C.~Parkes$^{*a}$,\\
K.~Akiba$^{b,c}$,M.~Artuso$^d$,J.~Borel$^e$,T.J.V.~Bowcock$^f$,\\
J.~Buytaert$^b$,P.~Collins$^b$,R.~Dumps$^b$,L.~Dwyer$^f$,\\
D.~Eckstein$^b$,L.~Eklund$^a$,M.~Ferro-Luzzi$^b$,R.~Frei$^e$,\\
G.~Haefeli$^e$,K.~Hennessy$^g$,T.~Huse$^f$,E.~Jans$^h$,\\
M.~John$^b$,T.J.~Ketel$^h$,A.~Keune$^h$,T.~La\v{s}tovi\v{c}ka$^b$,\\
R.~Mountain$^d$,N.~Neufeld$^b$,A.~Papadelis$^h$,S.~Stone$^d$,\\
T.~Szumlak$^a$,M.~Tobin$^f$,M.~Van~Beuzekom$^h$,A.~Van~Lysebetten$^h$,\\
T.W.~Versloot$^h$,H.~de~Vries$^h$,J.C.~Wang$^d$\\
\vspace{-2mm}\\ 
$^a$ University of Glasgow, Glasgow, United Kingdom\\
$^b$ CERN, Geneva, Switzerland\\
$^c$ Universidade Federal do Rio de Janeiro, Rio, Brazil\\
$^d$ Syracuse University, Syracuse, United States\\
$^e$ Ecole Polytechnique F\'ed\'erale de Lausanne, Lausanne, Switzerland\\
$^f$ University of Liverpool, Liverpool, United Kingdom\\
$^h$ NIKHEF, Amsterdam, Netherlands\\
$^g$ University College Dublin, Dublin, Ireland\\
{\it $^*$ corresponding authors}}

\begin{abstract}
LHCb is the dedicated heavy flavour experiment at the Large Hadron Collider at CERN. The partially assembled silicon vertex locator (VELO) of the LHCb experiment has been tested in a beam test. The data from this beam test have been used to determine the performance of the VELO alignment algorithm. The relative alignment of the two silicon sensors in a module and the relative alignment of the modules has been extracted. This alignment is shown to be accurate at a level of approximately $2~\mum$ and $0.1~\mrad$ for translations and rotations, respectively in the plane of the sensors. A single hit precision at normal track incidence of about $10~\mum$ is obtained for the sensors. The alignment of the system is shown to be stable at better than the $10~\mum$ level under air to vacuum pressure changes and mechanical movements of the assembled system.
\end{abstract}


\end{frontmatter}

\section{Introduction}

As part of the LHCb VELO~\cite{bib:ReOptTDR-03} detector commissioning, a beam test experiment with a partially equipped system was conducted. This paper reports on the results obtained when applying the VELO alignment algorithm to the data collected. The alignment of the LHCb VELO is particularly critical due to the high precision required from the VELO for the physics programme of the experiment, the on-line use of the VELO in finding displaced vertices in the trigger system, and the requirement that the detector is retracted and re-inserted between each fill of the LHC machine. The alignment procedure is described in Ref.~\cite{bib:NIM-07}.

An overview of the detector setup used during the beam test is provided in Section~\ref{sec:ACDC}. The performance of the alignment algorithm is assessed in Section~\ref{sec:alignment} through the analysis of track residuals and a comparison of the results with the metrology survey of the modules. 
Section~\ref{sec:performance} reports the impact of the alignment on the detector performance.
It covers the diagnosis of a cluster position reconstruction bias due to cable cross-talk, the analysis of the alignment effect on the reconstruction of track vertices, and the determination of the resolution of the sensors. 
As the detector will be operated in vacuum and will be retracted and re-inserted for each LHC fill~\cite{bib:LHCb-08}, the stability of the alignment under pressure variations and mechanical movements is reported in Section~\ref{sec:stability}. Section~\ref{sec:conc} summarizes the main conclusions.

\section{Beam Test Setup}
\label{sec:ACDC}

A partially equipped VELO detector half was tested in November 2006 in a $180~\GeVp$ hadron and muon beam at the CERN SPS. The mechanical suspension, cooling system and vacuum operation were designed to provide a good representation of the conditions expected from the final experiment. Ten of the 21 modules in one half of the detector were installed in their final setup. Each VELO module contains two approximately semi-circular silicon sensors: one with strips that are sectors of a circle, known as the \R\ sensor; and one with pseudo-radial strips, known as the $\Phi$ sensor. 
The strips on the $\Phi$ sensor are divided into an inner and outer sector with a positive and negative stereo angle, respectively.

\begin{figure}[h!]
\begin{center}
\setlength{\unitlength}{1cm}
\includegraphics[width=0.85\columnwidth]{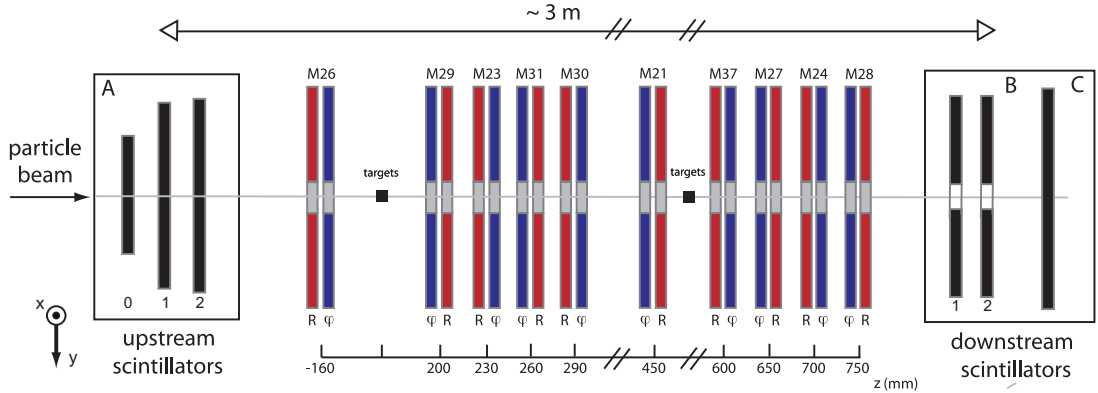}
\caption{Schematic top view of the beam test setup. A total of ten modules were mounted in the detector half. The module numbers are indicated and the location of the \R\ and $\Phi$ sensors in the modules. The location of the targets is also shown.}
\label{fig:layout}
\end{center}
\end{figure}

Six out of the ten installed modules were readout simultaneously. Data was taken with several different cabling configurations for the module readout. Particles were observed directly from the beam or from interactions of the beam with a series of targets. The 1~\mm\ radius 300~\mum\ thick circular lead targets were installed to represent the primary vertex location that will be obtained in the final experiment. Figure~\ref{fig:layout} shows a schematic overview of the mounted modules. The coordinate system used, as indicated, is equivalent to that used in the final LHCb detector.

The electronic read-out system and prototype data processing algorithms of the final experiment were applied. Analogue information is obtained from each channel and digitised to 10-bit precision. Pedestal subtraction and common mode suppression algorithms were applied. In addition it was found that it was necessary to remove cable cross-talk effects, see section~\ref{sec:FIR}, for which a Finite Impulse Response (FIR) filter was used. A clustering algorithm with a weighted pulse-height centre calculation was applied. In the final experiment these algorithms will be applied in firmware in field programmable gate arrays, here a bit-perfect C-emulation of these algorithms has been used. The detector half was operated under vacuum ($10^{-3}$~mbar) with modules cooled down ($<~0~^o$C). 

\section{Alignment Quality }
\label{sec:alignment}

\subsection{Alignment Algorithm}
\vspace{-2mm}
The LHCb VELO alignment algorithm is presented in Ref.~\cite{bib:NIM-07}. The alignment proceeds in three stages: the relative alignment of the \R\ and $\Phi$ sensors inside a module; the relative alignment of the modules inside a detector-half; and the relative alignment of the two detector-halves. The results of applying the first two stages of the procedure are reported here.

The relative alignment of the \R\ and $\Phi$ sensors inside a module is extracted from a fit to the distribution of track residuals as a function of the $\phi$ co-ordinate, that is from the characteristically curved shapes that are shown in Figure~\ref{fig:PhiBefore}.
The relative alignment of the modules inside a detector half is determined by a non-iterative matrix inversion technique. The residuals are expressed as a linear function of both the individual track parameters and the module alignment constants. The alignment constants are extracted through the inversion of only the components of this large matrix that involve these alignment constants. 

The results presented here used the data from two read-out cabling configurations and primarily use data in which the beam passed through the targets, as this contained a complimentary set of tracks both perpendicular and at small angles to the sensors.

The relative positions of the $\R$ and $\Phi$ sensors inside the individual modules and the relative position of the modules were initially assumed to be at their nominal design positions. Corresponding alignment constants were applied as the starting point for the alignment procedure. The software algorithms to determine the relative alignment of the \R\ and $\Phi$ sensors and the relative alignment of the modules were then applied and the results are presented in the following sections.

\subsection{Residual Distributions}

The observed signals on sensor strips are used to determine the best estimate of the hit cluster position. The hits on the sensors (other than those in the module under study) are fitted to produce tracks. These tracks are extrapolated to obtain the track intercept point in the sensor under study.  The unbiased residual is then defined as the perpendicular distance between the track intercept point and the line parallel to the strip at the observed cluster position. Consequently, these residuals are sensitive to sensor misalignments perpendicular to the strip direction.

The distribution of the residuals across the sensor surface is sensitive to misalignments. For example, as described in Ref.~\cite{bib:NIM-07}, plotting the $\Phi$ and $R$ sensor residuals as a function of the $\phi$ co-ordinate gives direct information on the relative $x$,$y$ translations of the sensors. In the case of a perfect alignment these distributions should be flat when plotted against any co-ordinate variable.

\begin{figure}[h!]
\begin{center}
\begin{minipage}[t]{6.5cm}
    \resizebox{6.5cm}{!}{\includegraphics{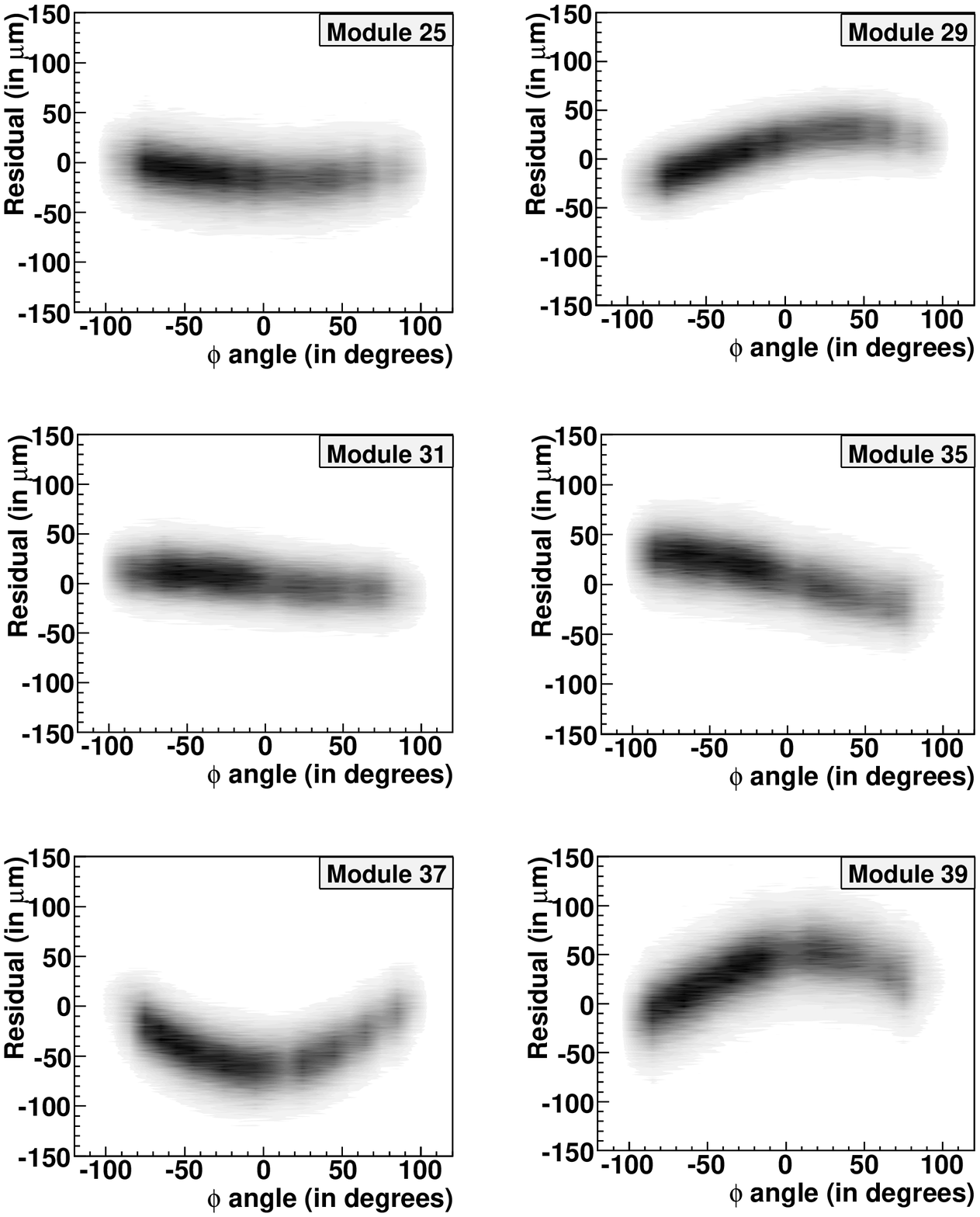}}
    \caption{Unbiased $\Phi$ sensor residuals without any alignment information as a function of the $\phi$ co-ordinate.}
    \label{fig:PhiBefore}
\end{minipage}
\hfill
\begin{minipage}[t]{6.5cm}
    \resizebox{6.5cm}{!}{\includegraphics{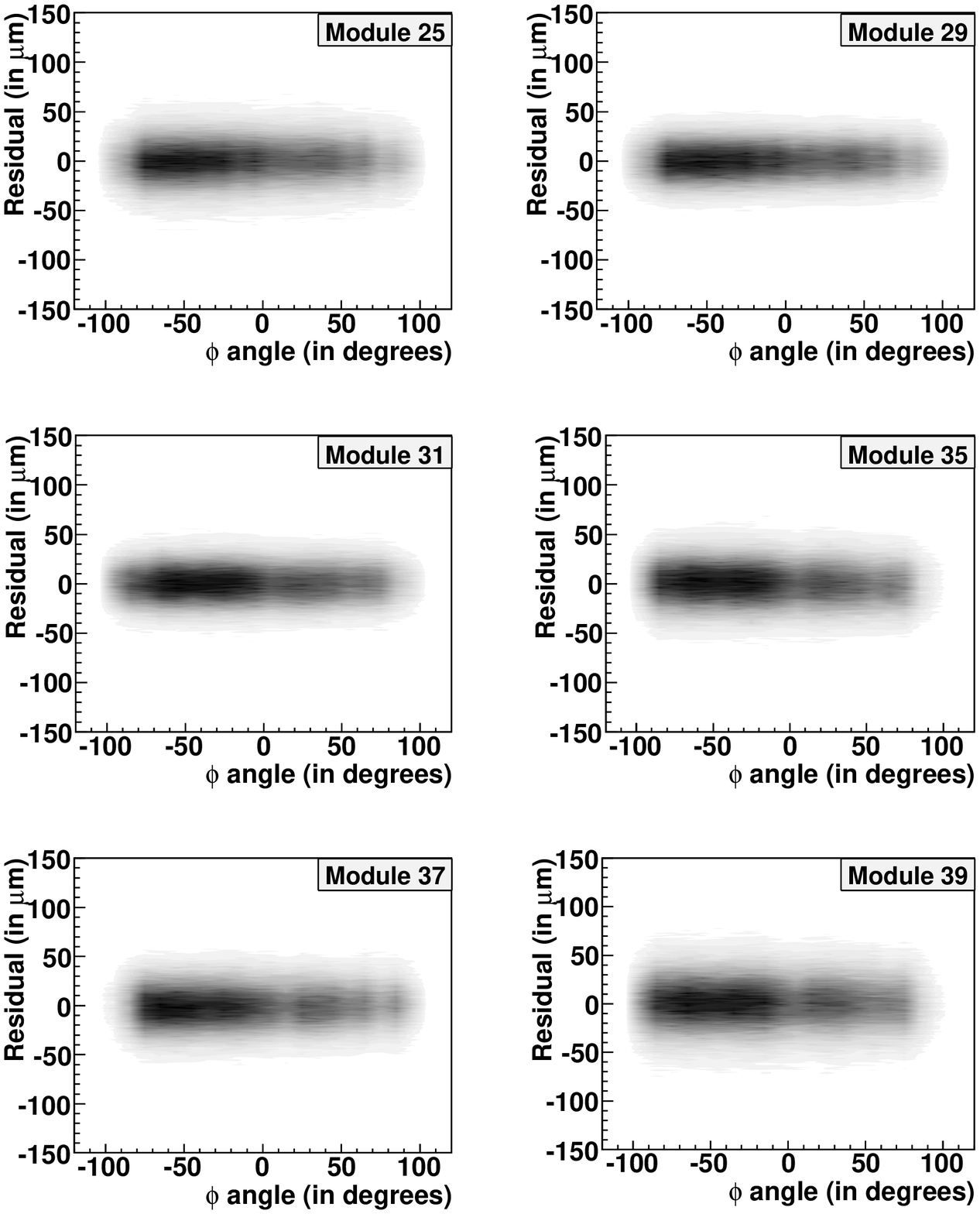}}
    \caption{Unbiased $\Phi$ sensor residuals after alignment (survey and software) as a function of the $\phi$ co-ordinate.}
    \label{fig:PhiAfter}
\end{minipage}
\end{center}
\end{figure}

Figure~\ref{fig:PhiBefore} shows the distribution of residuals on the $\Phi$ sensor plotted against the $\phi$ co-ordinate before the alignment procedure has been performed assuming the alignment constants are as in the nominal detector design. The results after applying the alignment procedure are shown in Figure~\ref{fig:PhiAfter}. Equivalent results were obtained for \R\ sensors. As expected, applying the alignment information results in reducing the deformations in the distributions (which result primarily from the $x$ and $y$ displacements of the sensors) and moving the mean of the residuals toward 0 (which is primarily an effect of rotations around the $z$ axis).

\begin{figure}[h!]
\begin{center}
    \resizebox{10cm}{!}{\includegraphics{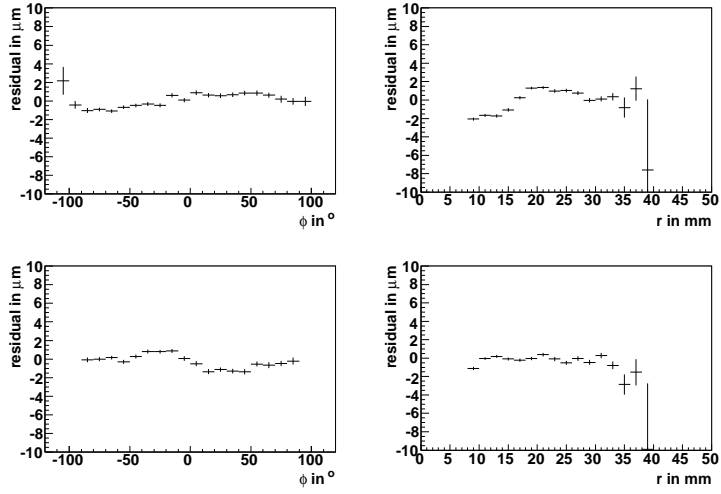}}
    \caption{The distribution of $\Phi$ (top) and $\R$ sensor (bottom) residuals in module 25 as a function of $\phi$ (left) and \rvar~co-ordinate (right).}
    \label{fig:singleModule}
\end{center}
\end{figure}

In Figure~\ref{fig:singleModule} the mean of the residual distributions for the $\Phi$ and $R$ sensors in one typical module are shown plotted against both \rvar\ and $\phi$ co-ordinates after the alignment procedure has been applied.
The distribution of residuals on the $\Phi$ sensor plotted against \rvar\ is seen to have a small change at the transition radius ($r=17~\mathrm{mm}$) between the inner and outer sectors of the sensor.
This effect is under study.

\begin{figure}[h!]
\begin{center}
\begin{minipage}[t]{6.5cm}
    \resizebox{6.5cm}{!}{\includegraphics{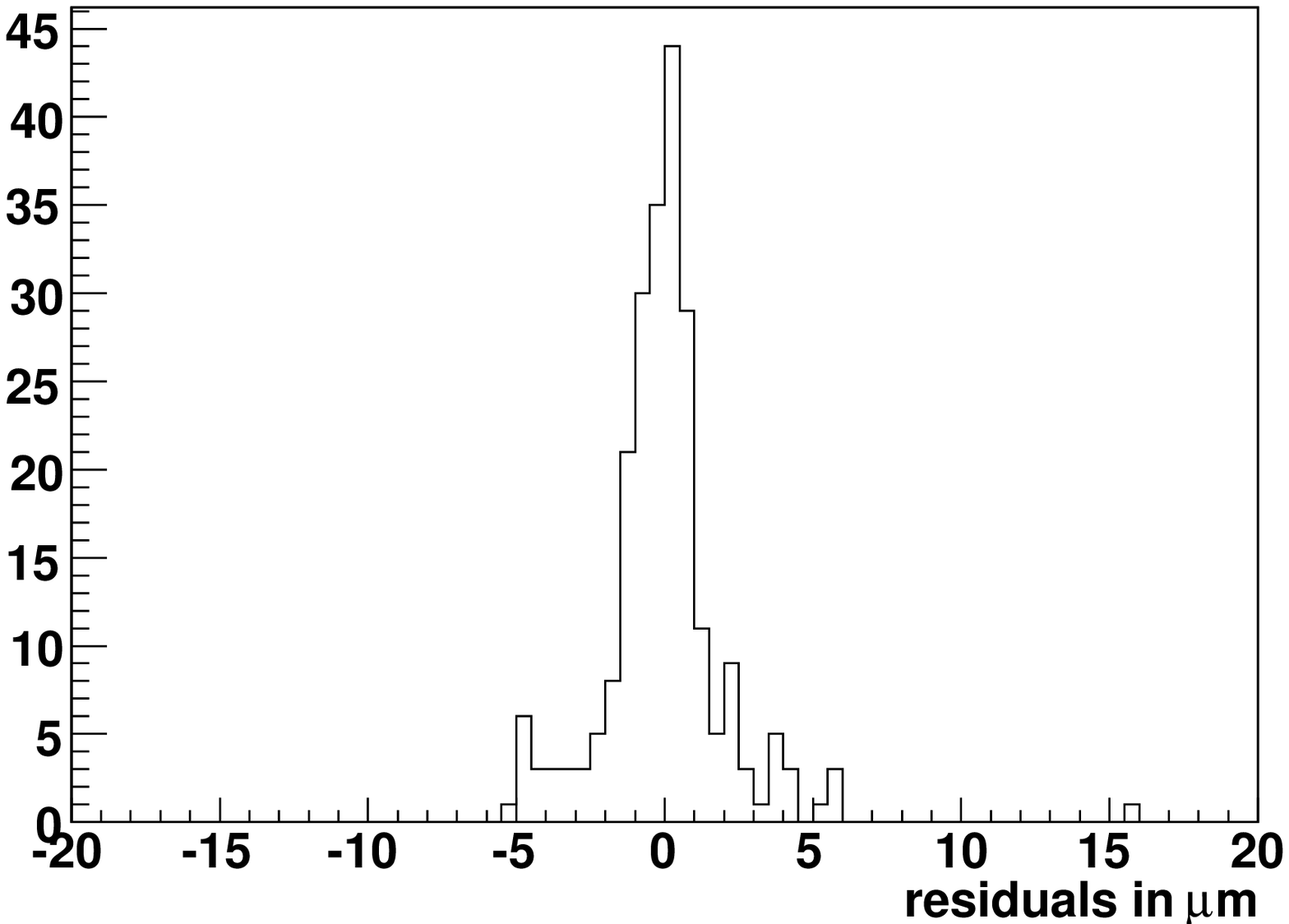}}
    \caption{Alignment precision of $x$ and $y$ translations.}
    \label{fig:TransQual}
\end{minipage}
\hfill
\begin{minipage}[t]{6.5cm}
    \resizebox{6.5cm}{!}{\includegraphics{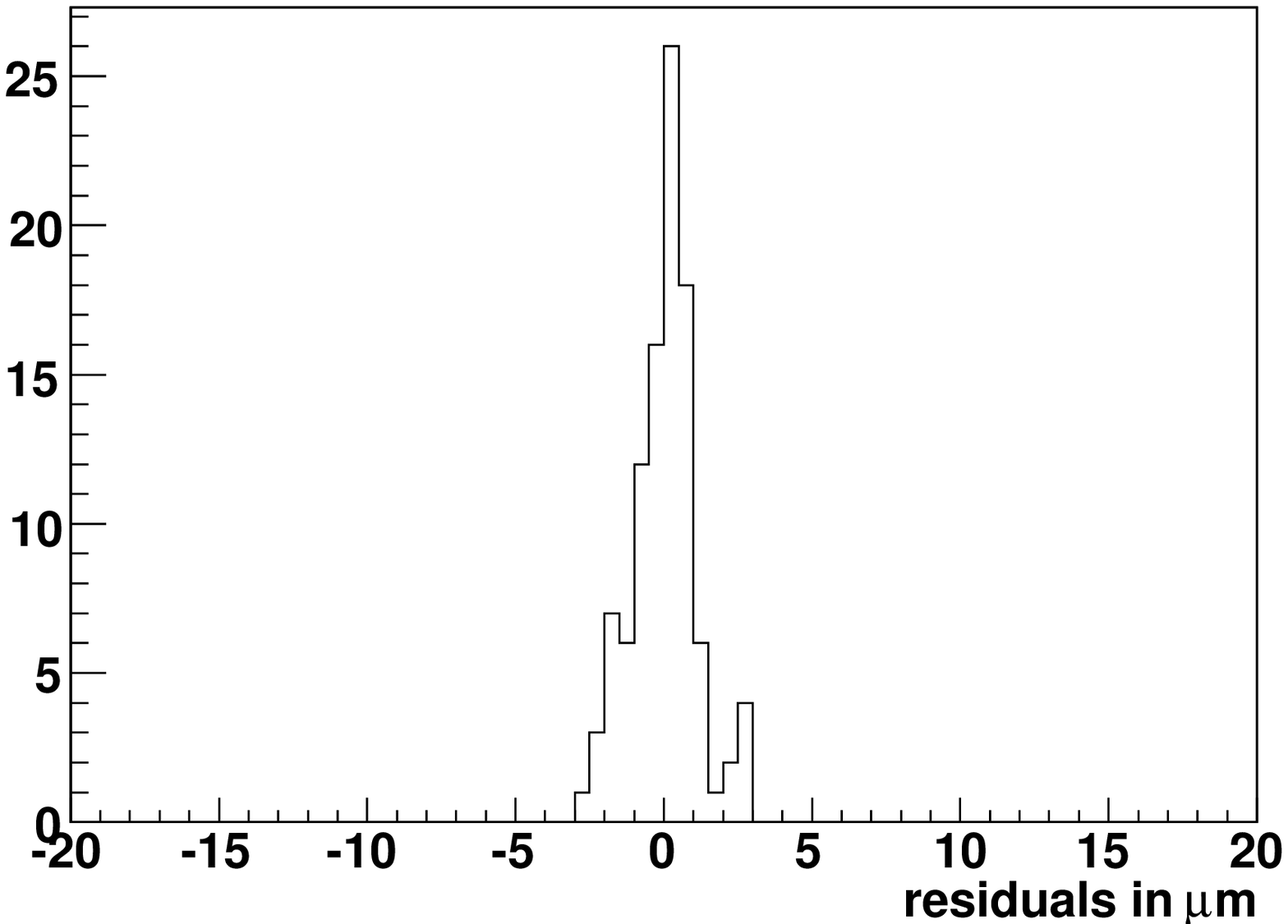}}
    \caption{Alignment precision of rotations around the $z$ axis.}
    \label{fig:RotQual}
\end{minipage}
\end{center}
\end{figure}

To assess the quality of the alignment achieved the mean of the residual distributions can be projected on the residual axis. The spread of this distribution then provides information on the remaining misalignments. These distributions obtained from all six sensors readout in a particular cabling configuration are shown in Figures~\ref{fig:TransQual} and~\ref{fig:RotQual}.

Figure~\ref{fig:TransQual} shows the projections of the residual means as a function of $\phi$ for all twelve sensors (six \R\ and six $\Phi$ sensors) under study. The r.m.s. of this distribution confirms that the $x$ and $y$ translations of all sensors are known to a precision of $2.1~\mum$, in good agreement with the combined precision of $1.1~\mum$ for the module alignment and $1.3~\mum$ for the sensor alignment as obtained from the simulation studies reported in Ref.~\cite{bib:NIM-07}.

Figure~\ref{fig:RotQual} shows the projections of the residual means as a function of $r$. This plot is primarily sensitive to rotations around the $z$ axis. The projection is made only for the data from the six $\Phi$ sensors since the $R$ sensors are insensitive to $z$ rotations. The measured r.m.s. of $1.1~\mum$ relates to the quality of constraining the rotations around the $z$ axis at an effective radius. This radius has been determined to be $11~\mm$ and hence these rotations are known to a precision of $0.1~\mrad$, matching that reported in Ref.~\cite{bib:NIM-07}.

\subsection{Comparison with Metrology}

\begin{figure}[h!]
\begin{center}
    \resizebox{6.5cm}{!}{\includegraphics{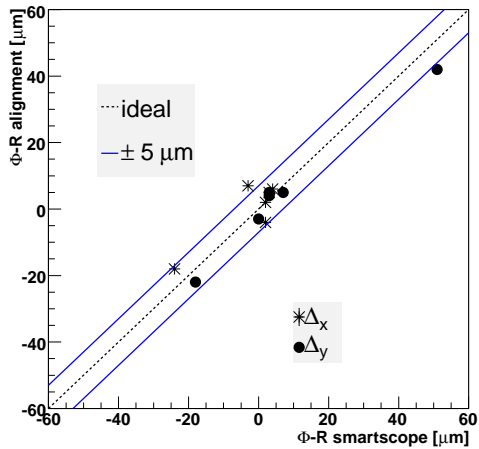}}
    \caption{Comparison of optical metrology results and software alignment results for relative sensor translations.}
    \label{fig:Sm_Ali}
\end{center}
\end{figure}

During the module production quality assurance tests, an optical metrology survey of the relative positions of the \R\ and $\Phi$ sensors inside the individual modules was performed. Another quality measure of the alignment can be obtained from the comparison of the alignment constants as determined by the software alignment with those measured by the metrology of the individual modules. The comparison between the two sets of measured constants for the relative sensor translations on each of the modules is shown in Figure~\ref{fig:Sm_Ali}. Agreement between both methods at a level of about $5~\mum$ is obtained, which is the expected precision of the metrology measurements.

\section{Effect of Alignment on Detector Performance}
\label{sec:performance}

This section reports on critical elements of the detector performance that are strongly affected by the alignment precision. The use of the alignment to diagnose a cluster centre reconstruction bias is discussed first. Then, a qualitative demonstration of the impact of the alignment on vertexing is shown. Finally, the sensor hit resolution after alignment is reported.

\subsection{FIR Filter}
\label{sec:FIR}

As a result of the high precision obtained from the detector alignment it has become possible to check for small scale effects in the detector geometry or biases in the cluster reconstruction position. A cluster centre reconstruction bias was observed and removed through the application of an FIR filter \cite{bib:Lars-07}. 

\begin{figure}[h!]
    \resizebox{6.5cm}{!}{\includegraphics{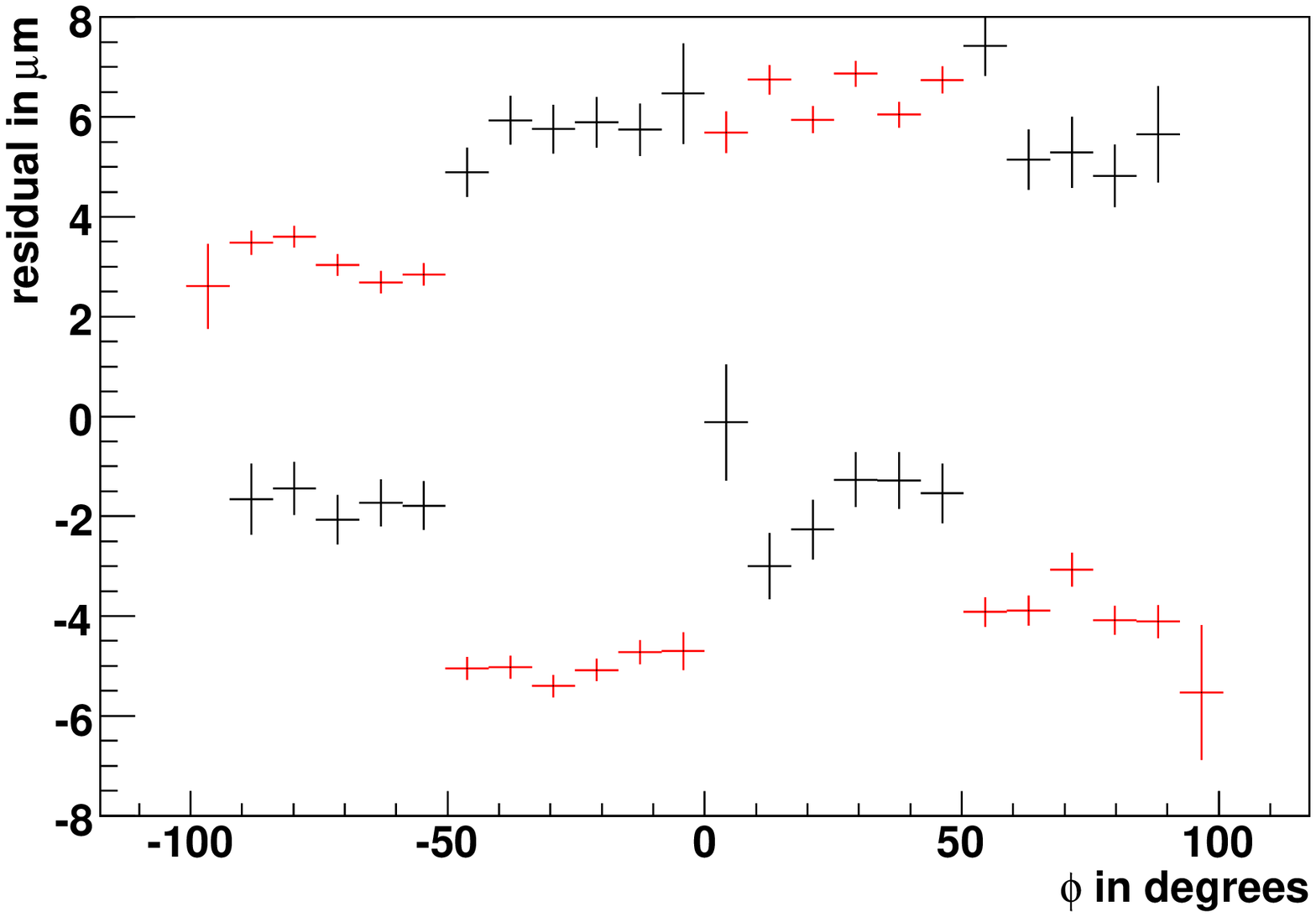}}
    \resizebox{6.5cm}{!}{\includegraphics{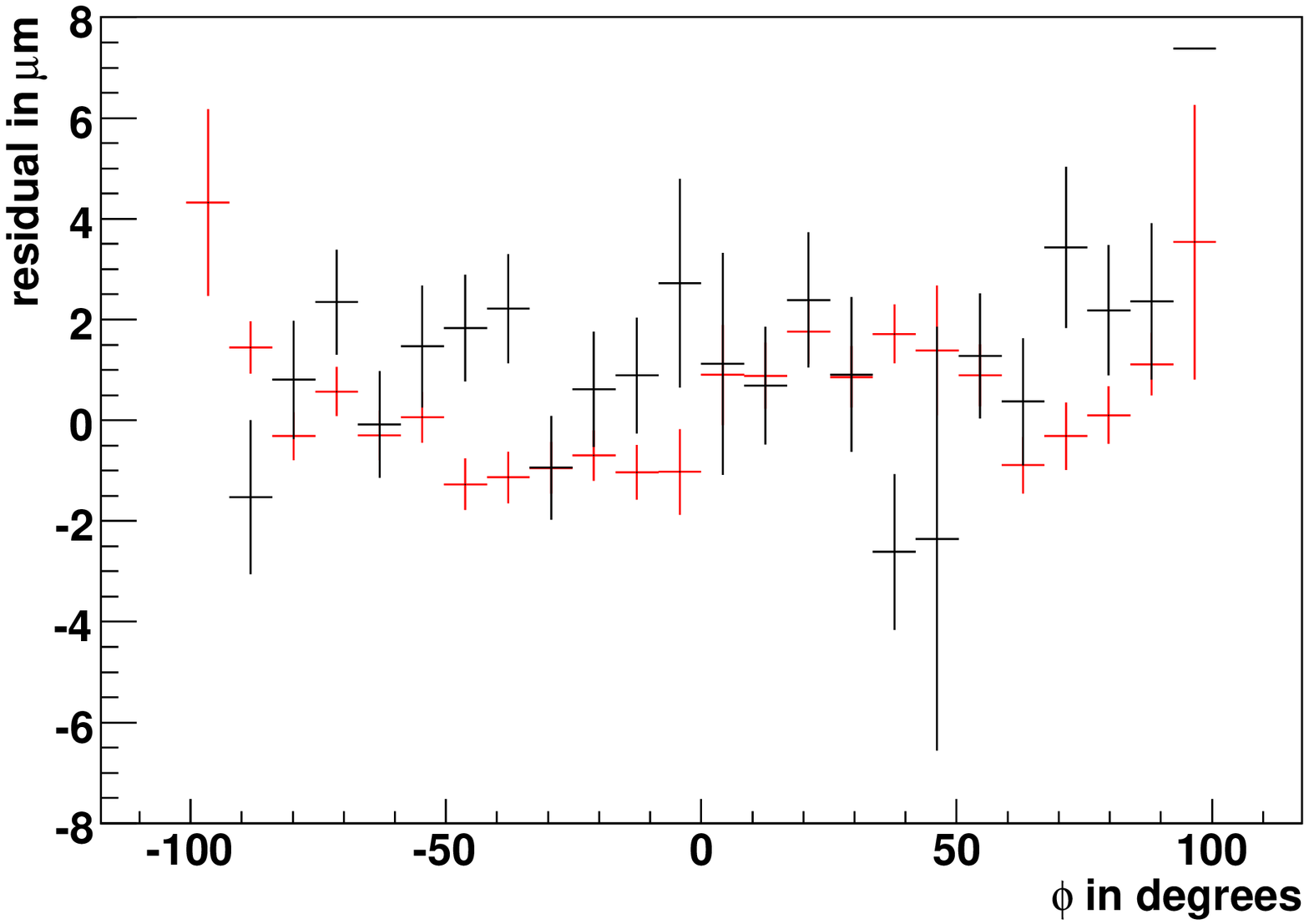}}
\caption{$R$ sensor residuals plotted against $\phi$ position across sensor, before and after an FIR filter is applied.
The residuals are split in groups of equal readout direction (see text).
In both cases the alignment procedure has been applied.
The right hand figure has reduced statistics.}
    \label{fig:FIR}
\end{figure}

The LHCb R-sensor contains four approximately $45^\circ$ sectors each with 512 strips. The strips are connected to bond pads at the outer edge of the sensor through the use of a double metal routing layer. In the first sector, the strips 127 to 0 are readout first, followed by those from 128 to 511. This pattern is reversed in adjacent sectors. Forward cross-talk between the analogue chip output signals in the cables to the readout board gave rise to a bias in the reconstructed signals and hence cluster positions. However, as a result of the readout pattern the direction of this bias reverses for sensor strips 0-127 and 128-511. The residual bias is clearly visible in Figure~\ref{fig:FIR} before the correction is applied. Once an FIR filter was applied to remove this effect the alignment quality was improved.
This correction will be included for the final experiment.
However, as this requires a time consuming data reprocessing this correction has not been included for the results given in the previous section and Figure~\ref{fig:FIR} after correction is shown with reduced statistics.

\subsection{Vertexing Performance}

The LHCb trigger is based on a precise separation of B hadron decay vertices from the primary interaction vertices. Hence, a very good vertexing accuracy is crucial for the experiment.

\begin{figure}[h!]
\begin{center}
\begin{minipage}[t]{6.5cm}
    \resizebox{6.5cm}{!}{\includegraphics{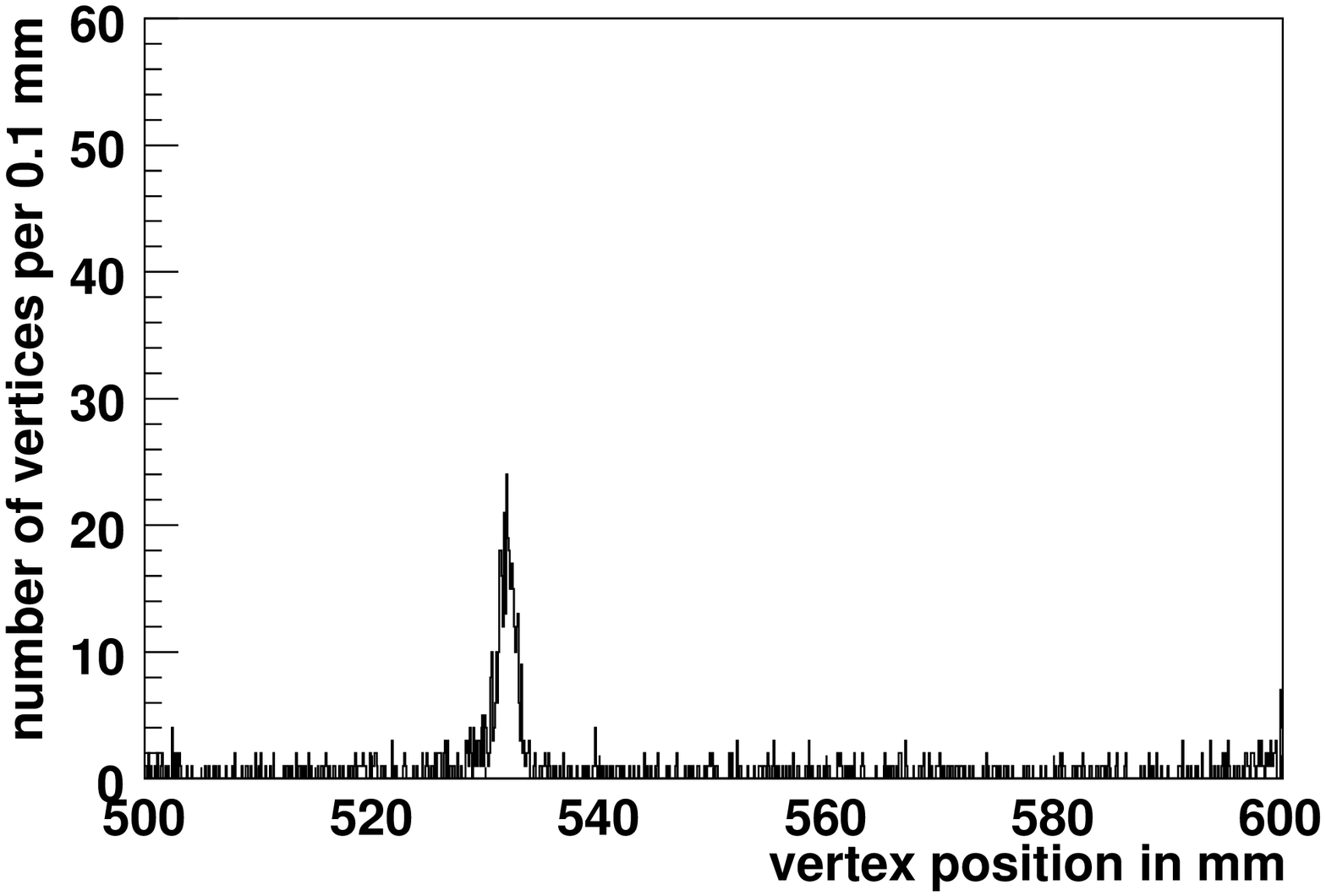}}
    \caption{Vertices reconstructed in targets before applying the alignment procedure.}
    \label{fig:V-Before}
\end{minipage}
\hfill
\begin{minipage}[t]{6.5cm}
    \resizebox{6.5cm}{!}{\includegraphics{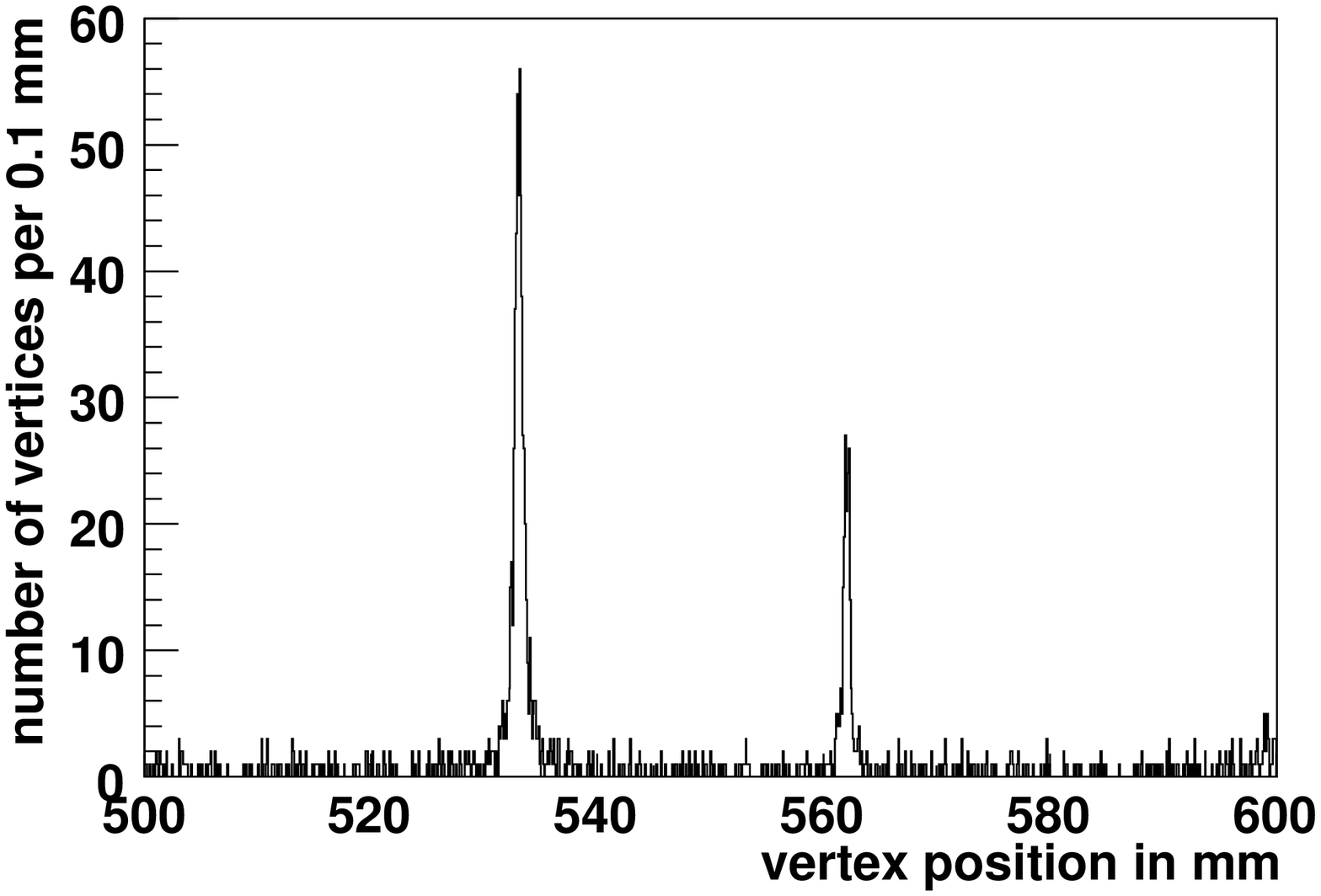}}
    \caption{Vertices reconstructed in targets after applying the alignment procedure.}
    \label{fig:V-After}
\end{minipage}
\end{center}
\end{figure}

The alignment quality of the VELO has a critical effect on the vertexing performance. This is demonstrated in Figures~\ref{fig:V-Before} and \ref{fig:V-After}. These plots were obtained on the same set of target events, using the same standard tuning for the pattern recognition. In Figure~\ref{fig:V-Before} the nominal design positions of the modules are applied, and it is apparent that only vertices from one of the two expected targets are observed. Figure~\ref{fig:V-After} shows the situation after the alignment procedure has been applied: the second target appears, and for the first target the precision of the vertex reconstruction improves significantly which also leads to a larger number of reconstructed vertices.

\subsection{Sensor Resolution}

The VELO sensor resolution has been determined using tracks of perpendicular incidence. The resolution has been determined from the sigma of a Gaussian fit to the distribution of the unbiased residuals. Both the \R\ and $\Phi$ sensors on the module under study were excluded in the track fit. The resolution is extracted as a function of the local strip pitch at the track intercept point. 

A correction has been applied for the error due to the extrapolation of the track to the sensor. The contribution was computed (see \cite{bib:LHCb-2000-099}) under the assumption that all sensors have equal performance and that the local pitch of all the \R\ or $\Phi$ sensors on a track is the same. This assumption is valid as the tracks angles are less than $2~\mrad$.

\begin{figure}[h!]
\begin{center}
    \resizebox{13cm}{!}{\includegraphics{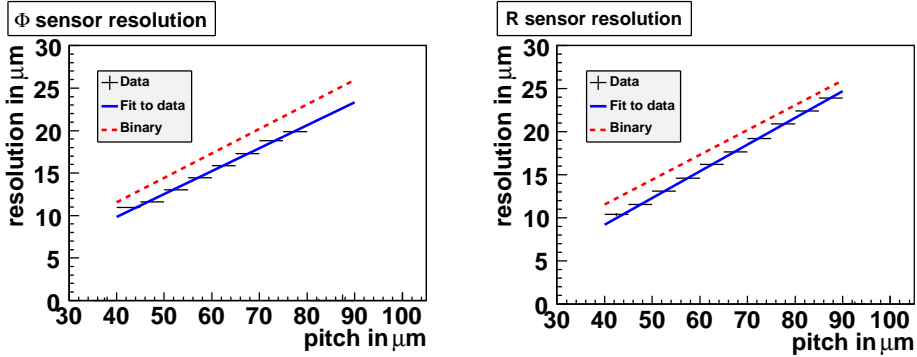}}
    \caption{$\Phi$ (left) and \R (right) sensor resolutions, averaged over 6 sensors.}
    \label{fig:ResAligned}
\end{center}
\end{figure}

Figure~\ref{fig:ResAligned} shows the VELO sensor resolution, averaged over 6 \R\ and $\Phi$ sensors. 
A single hit precision of roughly $9.5 + 0.3 \times (\mathrm{pitch}-40)~\mum$ is obtained for both \R~and $\Phi$ sensors, for normal incidence tracks.
A significantly better resolution is expected for tracks at angles around $8^{\circ}$ for which the charge sharing between adjacent strips is optimal.
The resolution improves significantly with track angle as the charge sharing is increased. The resolution has been extracted using a simple weighted pulse height algorithm for reconstruction of the cluster position. Additional development of the clustering algorithm is expected to further improve the precision.

\section{Alignment Stability}
\label{sec:stability}

The VELO modules are placed as close as possible to the LHC beam, to optimise the vertex reconstruction capability. As a result the sensors will be operated in vacuum separated from the primary vacuum of the LHC by an RF foil. Furthermore, due to the LHC beam stability during injection, the modules must be retracted by 30~\mm\ and reinserted for each fill. Hence, the stability of the alignment to pressure variations and mechanical movemements is of great importance.

\begin{figure}[h!]
\begin{center}
\begin{minipage}[t]{6.5cm}
    \resizebox{6.5cm}{!}{\includegraphics{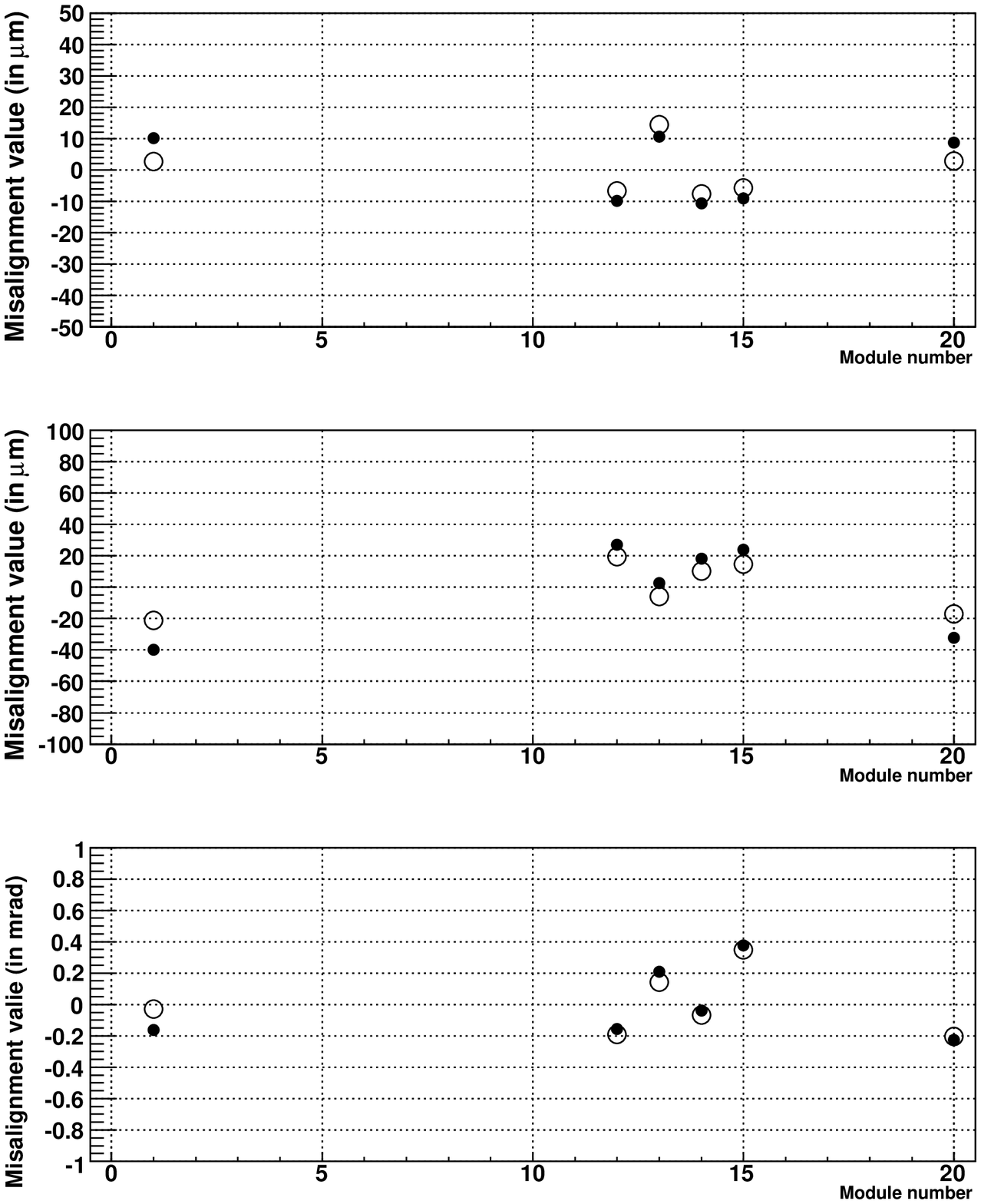}}
    \caption{Misalignment values in air ($\bullet$), and in vacuum ($\circ$).}
    \label{fig:Vacuum}
\end{minipage}
\hfill
\begin{minipage}[t]{6.5cm}
    \resizebox{6.5cm}{!}{\includegraphics{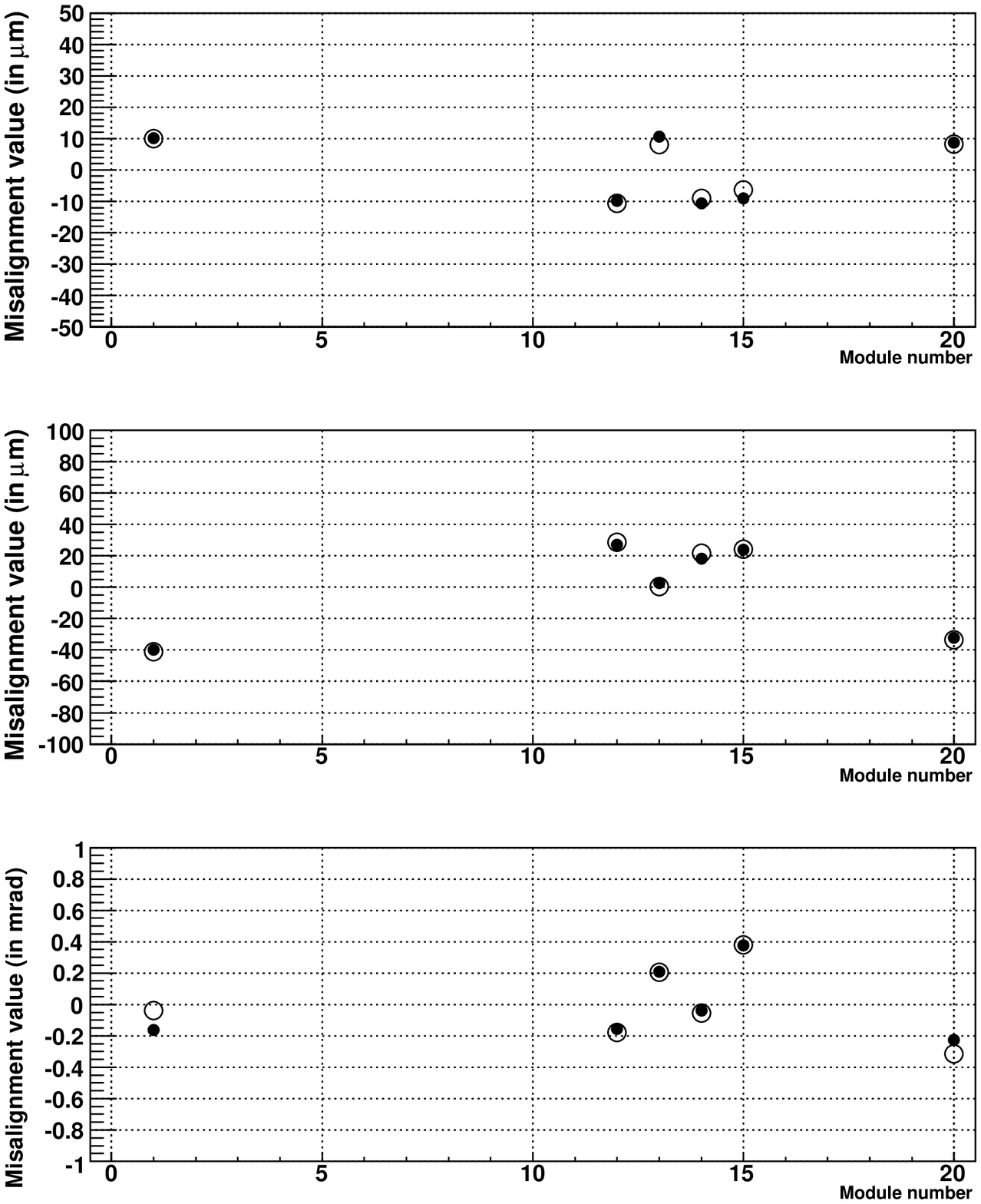}}
    \caption{Misalignment values before ($\circ$), and after ($\bullet$) detector halves movement.}
    \label{fig:Movement}
\end{minipage}
\end{center}
\end{figure}

Carbon fibre structures, such as the VELO module bases, may deform when pumped down in a vacuum chamber due to release of large quantities of humidity absorbed in the mesh.
Figure~\ref{fig:Vacuum} shows the alignment constants determined with data taken at atmospheric pressure and re-determined using data collected after the air was pumped out of the VELO vacuum vessel and a pressure of $10^{-3}$~mbar obtained. 
The three major degrees of freedom are shown: translations along x and y-axis (top and middle plots), and rotations around z-axis (bottom plot). These plots show that module movement as a result of the pumping operation are small ($<$~10 microns). This is an important result, particularly for the x translations, where the mechanical constraints are very tight for the distance between the sensors and their surrounding RF foil.

In order to move the beam from passing straight through the modules to hitting the targets, a shift of the experimental setup along the \x-axis was made, thus making a movement equivalent to the VELO half retraction, albeit with a different mechanical setup. 
The alignment constants before and after the movement are shown on Figure~\ref{fig:Movement}, using the same convention as for Figure~\ref{fig:Vacuum}. 
Again a very good stability of the module positions within their respective VELO half is observed. 
This result justifies the baseline assumption that the alignment procedure will not need to be performed on-line for each fill (for use in the trigger system). 
Instead, it is expected that the previously determined alignment constants for the modules can be used, while the VELO half alignment constants are updated to sufficient precision ($5-10~\mathrm{\mu{}m}$) by knowledge from mechanical sensors of the VELO movement system. 
The alignment constants can then be refined for use in the off-line physics analysis.

\section{Conclusion}
\label{sec:conc}

The software alignment procedure developed for the LHCb vertex locator was successfully tested on data from a beam test of the partially assembled detector system. 
The alignment accuracy obtained is $2.1$~\mum\ and $0.1$~mrad for translations and rotations respectively in the plane of the sensors.
Application of the alignment procedure significantly improves the tracking and vertexing performance of the detector. A single hit sensor resolution of $9.5$~\mum\ at $40$~\mum\ inter-strip pitch was obtained for tracks of normal incidence.  The alignment of the system was shown to be stable under movement of the detector and vacuum pumping.

\end{document}